# Pulsed rotating supersonic source used with merged molecular beams

L. Sheffield, M. Hickey, V. Krasovitskiy, K.D.D. Rathnayaka, I.F. Lyuksyutov[1], D. R. Herschbach

*Department of Physics and Astronomy, Texas A&M University, College Station, Texas 77843, USA*

We describe a pulsed rotating supersonic beam source, evolved from an ancestral device [M. Gupta and D. Herschbach, J. Phys. Chem. A **105**, 1626 (2001)]. The beam emerges from a nozzle near the tip of a hollow rotor which can be spun at high-speed to shift the molecular velocity distribution downward or upward over a wide range. Here we consider mostly the slowing mode. Introducing a pulsed gas inlet system, cryocooling, and a shutter gate eliminated the main handicap of the original device, in which continuous gas flow imposed high background pressure. The new version provides intense pulses, of duration 0.1-0.6 ms (depending on rotor speed) and containing ~$10^{12}$ molecules at lab speeds as low as 35 m/s and ~ $10^{15}$ molecules at 400 m/s. Beams of any molecule available as a gas can be slowed (or speeded); e.g., we have produced slow and fast beams of rare gases, $O_2$, $Cl_2$, $NO_2$, $NH_3$, and $SF_6$. For collision experiments, the ability to scan the beam speed by merely adjusting the rotor is especially advantageous when using two merged beams. By closely matching the beam speeds, very low *relative* collision energies can be attained without making either beam very slow.

## I. INTRODUCTION

The compelling frontier of cold (< 1 K) and ultracold (< 1 mK) gas-phase molecular physics, bristling with prospective applications and challenges, has been amply surveyed in recent evangelical reviews[1-5]. At present, the most effective experimental approach has been to induce formation of alkali dimer molecules from ultracold trapped alkali atoms by photoassociation or Feshbach resonances[6-10]. Over the past decade, however, much effort has been devoted to widening the chemical scope ("beyond the alkali age") by developing means to slow and cool preexisting molecules. Chief among these means are cooling by use of $^3$He as a buffer gas[11-14], which reaches 0.3 K; Stark deceleration of beams of polar molecules, using multiple stages of timed electric fields, which (depending on the molecule) can reduce translational energy well below 100 mK [15-17]; and Zeeman deceleration of magnetic atoms and molecules in analogous fashion[18-22]. Other methods, as yet less well developed or requiring unusual circumstances, include filtering slow polar molecules from an effusive or buffer-gas cooled source[23,24]; deceleration by optical fields[25-26]; reflection from a moving surface[27]; Stark slowing a nonpolar molecule, $H_2$, by exciting it to a Rydberg state[28]; cooling SrF with three lasers to exploit highly diagonal Franck-Condon factors[29]; "milking" collisions with special kinematic constraints[30-33]; or attaching molecules to superfluid helium nanodroplets[34].

---

[1] Author to whom correspondence should be addressed. Electronic mail: lyuksyutov@physics.tamu.edu

Here we consider a mechanical means to produce intense beams of slow (or fast) molecules, applicable to any substance available as a gas at ambient temperatures[35-37]. This employs a supersonic nozzle mounted near the tip of a high-speed rotor, which when spun contrary to the exiting beam markedly reduces the net lab velocity. An exploratory prototype device proved able to slow down, e.g., a Kr beam to below 42 m/s [35]. Recently, an improved version has been developed at Freiburg University[37]. It has a carbon fiber rotor, enhanced pumping, gas injection via a rotary feedthrough with ferrofluidic metal seals, and translation stages enabling adjustment of the nozzle position even during operation. Also, for use with polar molecules, the rotor source was augmented by an electrostatic quadrupole focusing field. We have developed another improved version of the rotating source. It does not include any of the Freiburg improvements, so is almost as rudimentary as the original version. The most important new feature is a pulsed gas inlet system, coupled with a gated shutter preceding the beam skimmer. This eliminated a major handicap of the original device (still present in the Freiburg version), in which continuous gas flow imposed high background pressure in the rotor chamber, both attenuating the yield of slow molecules and creating an interfering effusive flow into the detector chamber.

As well as providing slow beams with the familiar virtues of a supersonic molecular beam (high intensity, narrowed velocity distribution, drastic cooling of vibration and rotation), the rotating source enables scanning the lab beam velocity over a wide range by merely adjusting the rotor speed. For a stationary source shifting the beam velocity can only be done rather coarsely and awkwardly by changing the beam temperature or the ratio of seeded to carrier gas. A rotating source is subject to an intrinsic disability, however, because transverse spreading of the beam, which becomes more pronounced as the beam slows, causes the beam intensity to fall off with the square of the velocity[35-37]. That is a severe limitation; e.g., compared with a stationary supersonic source, the intensity of a Xe beam from the rotating source drops a hundredfold when the lab velocity is lowered to ~100 m/s [35]. Fortunately, this limitation can be avoided in pursuit of our prime goal, the study of slow collisions. The redeeming strategy makes use of merged beams, to obtain very low *relative* collision energies. Then neither beam needs to be particularly slow, provided the beam speeds can be closely matched. Merged beams have been extensively employed for ion-molecule, ion-ion, and electron-molecule collisions, using beams with keV energies to perform experiments at relative energies below 1 eV [38]. In similar fashion, merging a beam with fixed speed from a stationary source with the beam of adjustable speed from the rotor source can provide access to relative collision energies in the millikelvin range.

In keeping with customary practice in the field of cold (< 1 K) and ultracold (< 1 mK) molecules, we usually use degrees Kelvin (or milliKelvin) as the energy unit. 1 K = 3.16 millihartrees = 0.0862 meV = 0.695 cm$^{-1}$ = 1.98 cal/mol = 8.28 joule/mol.

## II. APPARATUS

Figures 1 and 2 show schematic views of our current apparatus, configured for merged-beam experiments. The stationary source (labeled **1**) supplies a pulsed supersonic beam of H (or D) atoms, produced by dissociation of H$_2$ (or D$_2$) in an RF discharge[39] and seeded in Xe or Kr carrier gas. The rotating source (labeled **2**) has the same basic anatomy pictured in ref.[35]; the mounting, balancing procedure, and AC induction driving motor are also the same. For the most part, we describe only differences in design and performance. The barrel of the new rotor was made of an aluminum alloy 7068 T6 (Kaiser Aluminum) with 35% higher yield strength than alloy 7075 T6 in ref.[35] and its length, from the axis of rotation to the nozzle exit aperture, was increased from 9.9 cm to 14.9 cm. The barrel with 1/8" ID is tapered in six steps; from thick to thin, the diameters of the six cylindrical segments are as follows: 0.625, 0.492, 0.412, 0.352, 0.300, 0.260 in. Their lengths are as follows: 3.231, 2.145, 1.100, 0.700, 0.600, 1.525 in. These dimensions, determined from computations analyzing the centrifugal forces[40], maximize the peripheral velocity at which the rotor should break; this theoretical limit is 633 m/s. The exit aperture, made from aluminum, located 2.5 mm from the tip of the barrel, was enlarged from a pinhole of diameter 0.1 mm to a conical shape of length 3 mm, cone angle 30°, and orifice diameter 0.4 mm.

The beam gas is fed into the spinning rotor via a stationary tube, which inserts into a steel sting whose inner diameter (1.65 mm) is slightly larger than the outer diameter of the feed tube. The sting (6.4 mm o.d.) is press-fitted into a hole in the topside of the rotor barrel, centered on the rotation axis. As in ref.[35], we used a feed tube (i.d. 0.75 mm) of PEEK (PolyEtherEtherKetone), which has flexibility and low friction similar to Teflon, but is more robust. The intrinsic leaking that occurs was rendered insignificant by adding a small differentially pumped auxiliary vacuum chamber to house the gas feed. As shown in Figure 3, the sting extends into the auxiliary chamber via a snug hole (6.5 mm dia) in a washer-shaped seal made of PEEK. There is very little leaking through this seal because the auxiliary chamber exhausts to a rotary pump. The PEEK tubing and washer need to be replaced periodically, after about 50 hours of operation. This input system has proved adequate to feed gas at pressures up to 1.5 bars without appreciably affecting the vacuum in the rotor chamber.

The rotating source, in its original version[35] and in that at Freiburg[37], emits the input gas in a continuous 360° spray, from which only a thin slice passes through a skimmer to become a collimated molecular beam. Such profligacy overburdens conventional pumping. As well as allowing deleteriously high background in the rotor and detector chambers, it lowers the tolerable level of input gas pressure and thereby the quality of the supersonic expansion. These drawbacks led us to introduce a pulsed valve in the rotor gas inlet (indicated by "**PGV**" in Figure 1) and a shutter (**3** in Figure 1) in front of the skimmer (**4** in Figure 1) that gives entry to the detector chamber. For PGV we have used either a standard Parker Series 9 valve for non-aggressive gases, or Parker Series 2 valve for corrosive gases like $NO_2$, controlled by IOTA-ONE Solenoid Valve Controller by Parker or by a custom made controller. The use of custom made controllers helps significantly to reduce the price, taking into account that we operate up to four valves and shutter. Our custom made controllers produce rectangular voltage pulses with adjustable amplitude up to 300 V, duration (0.1ms and up) and time delay. Design of such circuits is described in detail in the standard electronics handbook[41]. The time required to open Parker Series 9 can be as short as 0.1ms. The duration that the PGV is open can be adjusted, typically between 1 ms and 20 ms. The shutter, guarding entry to the skimmer, is also controlled by a custom made controller similar to those used to control the input valves. We designed two types: a solenoid device with open-close cycle as short as 2 ms duration, and a hard drive based device about tenfold faster; the latter is very useful for creating short beam pulses and monitoring time-of-flight. The rotor position is monitored by an induction proximity sensor, to provide a time-zero for control of the PGV and the shutter, as well as for time-of-flight measurements. As the inlet gas pulses are much longer than the rotational period of the source, the output that reaches the detector when the shutter is open is a sequence of gas pulses spaced by the rotational period. Closing the shutter in synchrony with the rotor enables isolating a single gas pulse. Figure 4 shows typical raw time-profile data for a krypton beam, (a) with the shutter open and (b) with the shutter operated to transmit only one of the sequences of pulses.

Installing the pulsed inlet valve and gated shutter system much enhanced operation of the rotating source. In our apparatus, the main chamber wherein the rotor resides is pumped by a 6000 liter/sec oil diffusion pump and the detector chamber by a 500 l/s pump, both backed by rotary vane pumps. When the gas input is shut off, the ambient pressure in the rotor chamber is about $2 \times 10^{-7}$ Torr and in the detector chamber $2 \times 10^{-8}$ Torr. The two chambers communicate only by the skimmer orifice (3 mm dia). Previously, in our apparatus as well as that of ref.[35], when the rotor was continuously spraying gas, if the input pressure into the rotor reached 100 Torr, the pressure in the rotor chamber rose to nearly $10^{-4}$ Torr and in the detector chamber to $5 \times 10^{-6}$ Torr. Such high background in the rotor chamber severely scatters slow molecules; e.g., it is estimated that more than 90% of Xe atoms slower than 70 m/s would be scattered from a beam while traveling the 10 cm from rotor to skimmer[35]. When operating in the pulsed mode, with the input pressure into the rotor as high as 1.5 bars, we find that the pressures in the rotor and detector chambers surge to about $5 \times 10^{-5}$ Torr and $2 \times 10^{-7}$ Torr, respectively, during the 20 ms – 40 ms "shooting" time that delivers the pulses. After the surge, the pressures, as indicated by ion gauges, subside within a few minutes to the pre-shot levels.

The detector chamber contains a Residual Gas Analyzer (RGA-100, Stanford Research System) to monitor the parent beams. The RGA is fitted with an electron multiplier and ion counter and, in one of several modes, can detect a selected species via a quadrupole mass-spectrometer with time resolution of about 0.01 ms. Signals from the RGA electron multiplier are amplified by a custom made current amplifier with filter and subsequently by a voltage amplifier. The design of the current amplifier is similar to that described in standard electronics handbooks[41], tailored to the parameters of our equipment. The amplifier output is processed by a digital acquisition card (PCI-DAS 4020/21, from Measurement and Computing). An analog isolation amplifier is employed between the voltage amplifier and DAQ card to isolate the measuring electronics from the computer. LabView software was used to control the RGA and acquire and average signals.

## III.  BEAM PROPERTIES

For stationary supersonic beam sources, principles and engineering practice are well established, both for continuous[42] and pulsed[43] versions. For rotating sources, the basic features seem to be the same (although as yet much less thoroughly examined). However, three distinctive features enter. The first is definitely advantageous, the second is compromised by pulsing the gas input, the third complicates time-of-flight analysis.

(*i*) As noted already, the rotating source enables scanning the beam velocity without changing the source temperature or seed-to-carrier gas ratio.

(*ii*) The rotating source acts as a gas centrifuge[35, 40]. For continuous input, this produces a density gradient that increases between the gas inlet at pressure $P_{in}$ on the rotation axis and the exit aperature at $R_{out}$. If the gas within the rotor remains at thermal equilibrium, the pressure behind the exit aperature, $P_0$, is governed by

$$P_0 = P_{in} \exp[mV_{rot}^2/2k_B T_0)] \tag{1}$$

where $V_{rot} = 2\pi\omega R_{out}$, with $\omega$ the angular velocity of the rotor, m the molecular mass, $k_B$ the Boltzmann constant, $T_0$ the source temperature. For continuous input, Eq.(1) is expected to be a fair approximation as long as $P_0 A_{out}$ is substantially less than $P_{in} A_{in}$, where the A's denote areas of the exit and inlet apertures. For our apparatus, that condition is satisfied for a wide range with $P_0 > P_{in}$ because $A_{out}/A_{in} = (4/3)^2 = 0.017$. For pulsed input, however, the gas flow through the rotor is inherently nonstationary. This is illustrated by data shown in Figure 5, displaying variations in the sequence of pulse amplitudes as the input pressure is raised from $P_{in} = 25$ to 454 Torr. Note that $P_{in}$ is the initial gas pressure in the mixing chamber (*cf.* Figure3) before the PGV is opened to release gas into the rotor. As $P_{in}$ is increased, the pulse with maximum amplitude occurs earlier in the sequence. This indicates that increasing $P_{in}$ boosts how quickly gas fills and drains from the rotor, eventually making the first pulse the largest. Particularly for high $P_{in}$, the rapid draining renders uncertain the pressure distribution within the rotor, so makes estimates of $P_0$ from Eq.(1) inapplicable. Although pulsing the gas input allows use of considerably higher $P_{in}$ than does continuous input, there remains the limitation imposed by formation of dimers and higher clusters. Criteria based on empirical results[42, 43] indicate that for Kr and Xe, the carrier gases we most use, $P_0$ should not exceed ~ 500 Torr, to keep dimerization below ~ 1%.

(*iii*) Because molecular beams have appreciable angular width, for a rotating source the range of "shooting positions" that allow beam molecules to pass through the skimmer is much broader than for a stationary source. Figure 6 indicates this range, which is determined by the ratio of the skimmer aperture to the length of the rotor between the gas inlet and the exit orifice. The range is $\phi_{max} = \pm 11.5°$ for our current rotor and skimmer set-up. Molecules emitted from the rotor at position $\phi = -11.5°$ travel further than those emitted at $\phi = + 11.5°$; the difference is $\Delta z = 2R_{out}\sin\phi_{max} = 6$ cm. The molecules emitted with different $\phi$ actually have the same spread in speed. In a TOF measurement, however, the disparity in travel distances introduces an apparent spread, $\Delta V_{app}$. That can become much larger than the actual spread, as $\Delta V_{app}/V$ is at least comparable to the ratio of $\Delta z$ to the distance between the nozzle exit aperture at the $\phi = 0$ position and the detector, which is 6/45 = 0.13 in our apparatus. Hence TOF data does not give realistic velocity spreads unless carefully deconvoluted[35-37]. For slower beams, formed at high rotation frequencies, TOF analysis is further complicated by "wrap-around" because then spreads in travel time become longer than intervals between pulses.

These complications and partial remedies are more fully discussed elsewhere, from the perspective of the original but conflicting aims of attaining beams both intense and slow enough to trap[35-37]. As now we intend to use the rotating source in merged beam collision experiments, slowing is no longer a major concern. Here we want to illustrate aspects most pertinent for the merging approach. These are the velocity scanning capability and how much the increase in input pressure enabled by pulsed operation can enhance the beam intensity and narrow its velocity spread. As a basis for assessment, we consider estimates obtained from standard approximate formulas for supersonic beams[42].

For the rotating source, the velocity distribution of molecular flux obtained on transforming into the laboratory frame[44, 35] is given by

$$F(V) = V^2(V - V_{rot})\exp\{-[(V - w)/\Delta v]^2\}, \tag{2}$$

aside from normalization; $V_{rot}$ is the peripheral velocity of the rotor, $w = u + V_{rot}$ is the flow velocity along the centerline of the beam in the laboratory frame, u the flow velocity relative to the rotating exit nozzle. In the slowing mode, when the rotor spins contrary to the beam exit flow, $V_{rot} < 0$; in the speeding mode, $V_{rot} > 0$. The velocity spread is governed by

$$\Delta v = (2k_B T_\parallel/m)^{1/2} \tag{3}$$

The parallel (also designated longitudinal) temperature of the expansion, $T_\parallel$, describes the molecular translational motion with respect to the flow velocity. According to the thermal conduction model[45], $T_\parallel/T_0$ is proportional to $(P_0d)^{-\beta}$, with d the nozzle diameter and the exponent $\beta = 6(\gamma - 1)/(\gamma + 2)$, with $\gamma = C_P/C_V$ the heat capacity ratio. Likewise, the flow velocity

$$u = (2k_BT_0/m)^{1/2}[\gamma/(\gamma-1)]^{1/2}[1 - (T_\parallel/T_0)]^{1/2} \qquad (4)$$

involves both $T_\parallel/T_0$ and $\gamma$.

The intensity delivered to the skimmer is proportional to $P_0d^2$. It can be determined by relating the integral of Eq.(2) over all velocities to the centerline intensity when the rotor is stationary, which can be obtained from familiar expressions[42] involving as well $T_0$, m, $\gamma$, and apparatus geometry as described in ref.[35, 36]. The calculated total beam intensity for $P_0d \sim 1$ is of the order of $10^{18}$ molecules/sr/s, a typical magnitude. Our RGA calibration proved extremely fickle, so did not provide a satisfactory confirmation. However, comparisons with experimental results of Ref.[35] are consistent with the calculated intensity. The corresponding estimated intensity per pulse, again for $P_0d \sim 1$, is about $10^{15}$ molecules. Figure 7 displays the variation with $V_{rot}$ and $P_0d^2$ of the number of molecules/sec predicted to arrive at the observation zone. Data points with curve (a) were obtained with our pulsed source; for (b) from[35], for (c) from[37]. The shape of the curves, calculated as described in Ref.[35], is governed chiefly by $V_{rot}$. To illustrate the dependence on $P_0d^2$, we simply shifted the curves for (a) and (c) relative to (b), which is taken from Figure 10 of Ref.[35]. Even for the Xe beam of curve (b), the centrifugal effect of Eq.(1) made only a minor contribution; as it would be much smaller for the Kr beam of (a) and Ar beam of (c), in Figure 7 we have omitted it for all three curves. Thus, we took $P_0 = P_{in}$. For the slowing mode the data points droop below the calculated curves, increasingly so as the lab velocity decreases ($V_{rot}$ more negative). For (b), a correction for attenuation by scattering by background gas, which becomes much more serious for slow molecules, was applied (open points). For (a) and (c), the background pressure is at least tenfold lower than in (b), and the droop is much less pronounced, but suggests some attenuation may still occur.

Figure 8 shows velocity distributions obtained from Eqs.(2-4) to illustrate that the width $\Delta v$ narrows as $P_0d$ is increased. That can occur either by increasing $P_{in}$ or by the centrifugal enhancement given by Eq.(1). When operative, the centrifugal effect can decrease the parallel temperature $T_\parallel$ below that for a stationary source by tenfold or more, so narrow $\Delta v$ more than threefold, as confirmed by experimental data presented in Ref.[35]. The relation between $T_\parallel$ and $P_0d$ given by the thermal conduction model has likewise been confirmed for rotating sources, both in Ref.[35] and Ref.[37]. Another aspect exhibited in Figure 8 is that a decrease in the Poisson ratio, $\gamma = C_p/C_v$, results in less efficient cooling as the backing pressure is increased. This is evident in comparing the widths for Kr ($\gamma = 5/3$) and $NO_2$ ($\gamma = 1.282$). Often "inverse" seeding (light seed, heavy carrier), is used to slow supersonic beams. For light molecules, that is done even with a rotating source to lower the range of $V_{rot}$ required. We note this because $T_\parallel$ should be lower for the light seed molecule than the carrier gas, according to theory[42] found consistent with experimental results[35, 36]. This offers a means, e.g. by seeding $NO_2$ in Xe, to offset the penalty imposed on $\Delta v$ by a smaller value of the Poisson ratio. The cost of that strategy, however, is a much lower centerline intensity of the seed gas, because of its small mole fraction and mass defocusing. We have adjusted the peak positions in Figure 8 to 420 m/s, to illustrate in (a) the upward effect of "slip" for H atoms seeded in Kr, and in (b) the downward shift provided by the counter-rotating source.

## IV. MERGED-BEAMS FOR SLOW COLLISION EXPERIMENTS

Returning to Figures 1 and 2, which depict a merged-beam experiment underway in our laboratory, we provide some details that serve to illustrate characteristic aspects. The distance from the stationary pulsed beam source **1** to the observation zone (OZ) is ~23 cm, and that from our pulsed counter-rotating source **2** at the rotor exit when in the nominal "shooting" position is ~13 cm. Our specimen reaction is $H + NO_2 \rightarrow OH + NO$. It has been much studied, both in "warm" beams[46-53] as well as other kinetic experiments and theory[54]. The beam **1** reactant is atomic H (or D), seeded in Kr; allowing for "slip" in the supersonic expansion[42], we estimate the most probable beam velocity is $V_1 \sim 420$ m/s, spread $\Delta v_1 \sim 35$ m/s. The beam **2** reactant is $NO_2$ (without carrier gas); by adjusting the rotor speed (which can be done to within ~ 1 m/s), we can obtain the same velocity, $V_2 \sim 420$ m/s, with estimated spread $\Delta v_2 \sim 50$ m/s. Use of the rotating source for $NO_2$ bestows an incidental bonus. In order to avoid appreciable dimerization to form $N_2O_4$, it is necessary to keep the gas within the source warm and the pressure modest (e.g., 300 K or above, 2 Torr or less). That is a severe constraint for experiments that require the ability to shift the velocity distribution substantially. Simply adjusting the rotor speed enables large shifts in the lab velocity of the beam with no change in the temperature and pressure within the source.

The transit time from source to OZ is considerably longer for beam **1** (nominal $t_1 = 0.66$ ms, spread 0.60-0.73 ms) than for beam **2** (nominal $t_2 = 0.37$ ms, spread 0.33-0.43 ms). Accordingly, the H beam exit valve is opened about 0.3 ms before the rotor reaches its "shooting" position, to ensure that H and $NO_2$ traveling at the nominal velocity arrive at about the same time at the OZ. High precision in the timing is not required. Complete spatial overlapping of the reactant beams in the OZ actually results simply because the pulse durations of both beams are long enough to allow long streams of molecules to issue forth. The H exit value is open longer than 1 ms, so the emitted beam pulse extends beyond 35 cm; likewise, at the rotor speed needed to produce a lab beam velocity of ~420 m/s, the pulse of $NO_2$ sent through the skimmer lasts longer than ~0.6 ms, so extends beyond 20 cm.

From the top view (Figure 1), it would appear that a traffic problem occurs, since beam **1** would intersect the rotor when it reached the "shooting" position. However, as shown in the side view depicted in Figure 2, beam **1** actually passes below the rotor (by about 3 mm). The rotor tip is made quite small (it only has to house the 0.4mm exit orifice), so there is sufficient clearance to ensure that only minor scattering occurs from the upper edge of beam **1** as it passes under the rotor. Such scattering is insignificant because only reactive collisions occurring in the OZ are detected.

The small angular spread imposed by the skimmer limits the intensity of the beams arriving at the OZ, although there is some compensation because the reactant beams merge in a pencil-like volume rather than cross perpendicularly. We have estimated from the pressures within the beam sources, exit and skimmer orifice diameters, and distances from the OZ, that in the OZ the density of our H beam is $n_1 \sim 10^{11}$ cm$^{-3}$ and that of our $NO_2$ beam $n_2 \sim 10^{12}$ cm$^{-3}$. Estimates for a typical "warm" crossed-beam study, obtained in the same way (as reported in Ref.[47]), are of the same order as ours for H but much larger for $NO_2$. If it proves necessary, by replacing our current rotor (radius 15 cm) by a smaller one (radius 5 cm), we could shrink distances sufficiently to increase $n_1$ nearly threefold and $n_2$ about tenfold at the OZ. The overall rate of formation of OH + NO (in various vibrational and rotational states) is given by $n_1 n_2 <k(E_R)> = n_1 n_2 <V_R \sigma(E_R)>$, where $k(E_R)$ is the reaction rate coefficient, $V_R$ the relative collision velocity, and $\sigma(E_R)$ the total reaction cross section. Brackets $<...>$ indicate an average over both internal states and the spread in $E_R$. We will detect OH by means of laser-induced fluorescence (LIF), as in warm experiments[47-50, 53].

The extensive application of merged-beams in high-energy experiments emphasizes and documents a favorable kinematic effect. It has enabled well-collimated beams with keV lab energies to be merged over interaction distances of tens of cm to study collisions at relative kinetic energies below 1 eV with high resolution[38]. In the center-of-mass system, contributions to the relative kinetic energy from the spreads in lab speeds of the beam particles are markedly "deamplified" by a kinematic factor proportional to the *difference* in the most probable lab beam speeds. This is readily demonstrated when the spreads are small fractions of the most probable speeds for both beams. For beams with lab speeds $V_1$ and $V_2$ intersecting at an angle $\theta$, the relative kinetic energy is

$$E_R = \tfrac{1}{2}\mu(V_1^2 + V_2^2 - 2V_1V_2\cos\theta) \qquad (5)$$

with $\mu = m_1m_2/(m_1 + m_2)$ the reduced mass. For perfectly merged beams, with $V_1 = V_2$ and $\theta = 0$, the relative kinetic energy would be zero (and no collisions could occur). If the spreads in speeds, $\Delta v_1$ and $\Delta v_2$ are very small, in first-order their contributions to $E_R$ are simply proportional to $|V_1 - V_2|$ and thus strongly deamplified when the most probable beam speeds, $V_1$ and $V_2$, are nearly equal. That occurs in many applications using high-energy beams, as the fractional spreads, $\Delta v/V$, are often only 0.1% or less[38]. For supersonic molecular beams, these spreads are typically 10% or more, so deamplification is less dramatic yet still pronounced.

We have computed the average relative kinetic energy, $<E_R>$, for merged supersonic beams with fixed $\theta$, using for both beams the flux velocity distribution for a stationary source, given by $F(V)$ of Eq.(2) with $V_{rot} = 0$ and $w = u$. The average is obtained in explicit form,

$$<E_R> = \tfrac{1}{2}\mu[u_1^2 f(x_1) + u_2^2 f(x_2) - 2u_1u_2 g(x_1)g(x_2)\cos\theta] \qquad (6)$$

with $x_i = \Delta v_i/u_i$ (i = 1,2) the ratio of velocity spread to flow velocity, specified by the ratio of Eqs. (3) and (4). The functions $f(x)$ and $g(x)$ are ratios of polynominal form with x-dependent coefficients:

$$f(x) = <(V/u)^2> = P_5(x)/P_3(x) \qquad (7)$$

$$g(x) = <(V/u)> = P_4(x)/P_3(x) \qquad (8)$$

where

$$P_n(x) = \sum_{s=0}^{n} c_s x^s \qquad (9)$$

and $c_s = \binom{n}{s} \Gamma(\{s+1\}/2, 1/x^2)$, comprised of binomial coefficients weighted by incomplete Gamma functions.

A fuller discussion of merged-beam distributions is given elsewhere[56], treating further a pair of stationary supersonic beams and also the case of one stationary beam, the other beam from a rotating source with the velocity distribution of Eq.(2). The full distributions, $P(E_R)$, are evaluated, as well as the rms energy spread, $\Delta E_R = [<E^2> - <E>^2]^{1/2}$. When the beam velocities are closely matched ($u_1 \approx u_2$ or $u_1 \approx w_2$) the form of $P(E_R)$ is qualitatively Poissonian, whereas if the velocities become more and more unmatched $P(E_R)$ becomes approximately Gaussian and then Maxwellian. Although for closely matched beams $<E_R>$ is minimal, the energy spread then reaches $\Delta E_R = 2^{1/2} <E_R>$, its maximal value. For modest unmatching, $<E_R>$ increases slowly while $\Delta E_R$ shrinks more rapidly.

For the conditions anticipated in our current H + $NO_2$ experiments, $u_1 = u_2 \sim 420$ m/s, $\Delta v_1 \sim 35$ m/s, $\Delta v_2 \sim 50$ m/s, and $\theta \sim 1.5°$, we find $<E_R> \sim 110$ mK. That is well within the "cold" collision realm (< 1 K), although the kinetic energies of the beams are 11 K and 490 K. In $<E_R>$, the velocity spreads contribute about 95%; if those were each reduced by 10 m/s, the relative kinetic energy would drop to ~65 mK. Figure 9 shows that $<E_R>$ for the $u_1 = u_2$ case can also be reduced substantially by lowering the matched flow velocity. Figure 10 shows how $<E_R>$ varies for modest mismatching of the flow velocities. Even a mismatch of ±10% will appreciably increase $<E_R>$ when both beams have similar velocity spreads, whereas a small mismatch becomes optimal when the velocity spreads differ considerably. Both Figs. 9 and 10 include curves for four sets of velocity spreads. These range from (a) utopian 1%, achievable for Stark decelerated beams[17], but at great cost in intensity; to (b) 5%, attainable with a compact velocity selector[55] with acceptable cost in intensity; to (c) 10%, typical for supersonic beams; to (d) 10% + 20%, similar to our current experiment.

## V. DISCUSSION AND OUTLOOK

To pursue gas phase "cold chemistry" in the mK range, the prime experimental requisite is sufficient flux of reactant collisions with very low *relative* kinetic energy. That is difficult to attain using either trapped reactants or crossed molecular beams, because then *both* reactants must contribute adequate flux with very low translational energies. Using merged-beams with nearly the same velocity can provide much higher flux with low relative energy because *neither* beam needs to be slow; instead, both can be operated in the usual warm range or not far below it. A rotating supersonic source can readily adjust its beam velocity over a wide range to match that from a stationary partner source. Pulsing both sources enables use of higher input pressures and thereby enhances the beam intensities. Also, gaining freedom from the need to produce slow but intense beams much widens chemical scope. The rotor source is suitable for any fairly volatile and docile molecule, whereas the stationary partner provides a complementary capability to generate species that must be produced from precursors, such as hydrogen, oxygen, or halogen atoms or free radicals. E.g., with little change, our current apparatus can be used for many reactions of H atoms, including with halogen or halogen halide molecules. With merged beams, candidate reactant molecules need not have properties amenable to manipulation, such as electric or magnetic moments, but it is advantageous to pair reactants that differ greatly in mass, since the relative kinetic energy is proportional to the reduced mass.

In merged-beam reactive collision experiments, the chief observable properties are the total cross section and its dependence on the relative kinetic energy of the reactants. These may be augmented by preparing internal states (electronic, vibrational, rotational) of the reactants or subjecting them to external fields. Merged-beams are not suited to observing the angular distribution or translational energy of reaction products. However, this situation accords with intrinsic limitations that enter in the slow collision realm[57]. There product angular distributions tend to become isotropic (when s-wave collisions predominate) or nearly so. Also, since reactions accessible in the cold realm are generally exoergic, disposal of energy and angular momentum among product states is virtually the same in cold collisions as in warm collisions. Hence, in cold reactive collisions, usually only reactant interactions can provide new information beyond that better found from experiments in the warm realm.

Another inviting aspect of merged-beam experiments is that they require mostly familiar molecular beam apparatus, not unusually expensive or virtuosic. Although a rotating source is uncommon, our current device has proved simple to assemble and robust in operation. It provides an especially convenient means to match velocities of the reactant beams. Surprisingly, we have found only three previous suggestions, all merely *en passant*, to apply merged beams to study low-energy collisions of uncharged molecules[35, 58, 59]. Prospects for merged-beams as a route to cold chemistry seem now to deserve more attention.

**Acknowledgements**

We are grateful for support of this work by the National Science Foundation (under grant CHE-0809651), by the Norman Hackerman Advamced Research Program (under grant 010366-0039-2007) and by the Robert A. Welch Foundation (under grant A-1688 to Lyuksyutov). Sheffield is a Welch Graduate Fellow. We are also thankful for support provided by Don Naugle, David Lee and Vladimir Khmelenko.We have enjoyed correspondence with Ronald Phaneuf about merged-beams as well as discussions with Manish Gupta about his pioneering rotating beam source. We have benefited from a design by Bretislav Friedrich of a prototype for the differentially pumped gas input (akin to our Figure 3), as well as efforts by Michael Timko to add a pulsed gas feed, and advice from Simon North and Mark Raizen on other experimental aspects. We thank Qi Wei for preparing Figures 7-10.

**Figures and Captions**

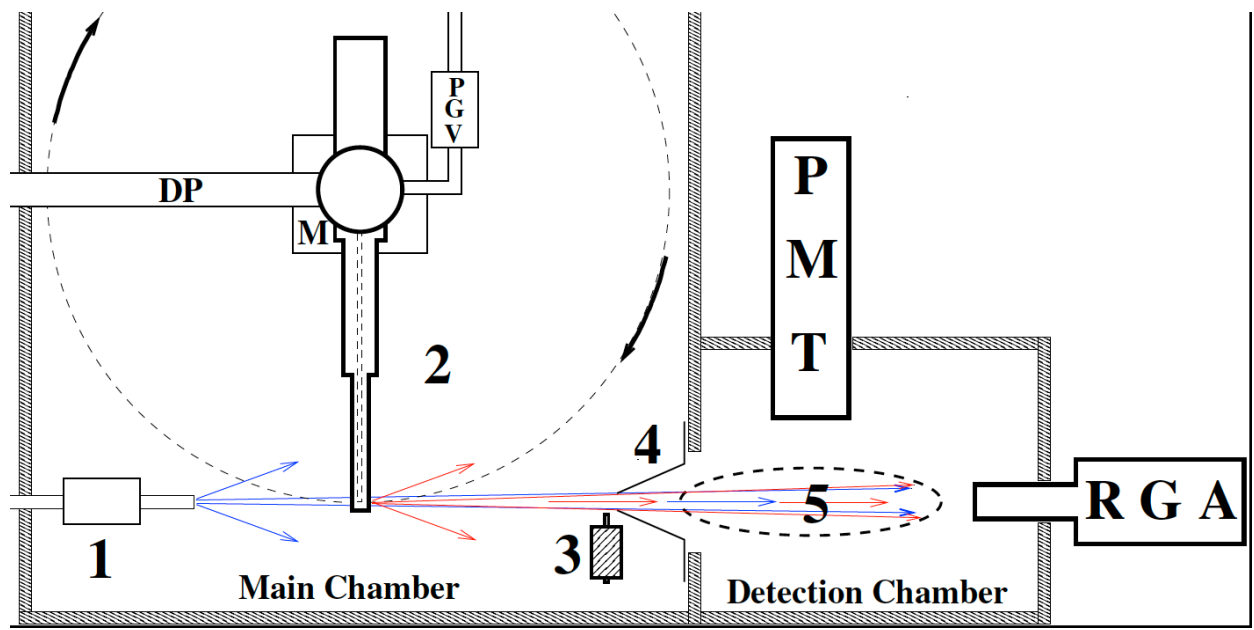

FIG. 1. Schematic (top view, not to scale) of basic apparatus, set-up in merged-beam mode, for study of H + NO$_2$ reaction: **1**) Stationary pulsed supersonic beam source of H or D atoms, formed in RF discharge (mounted outside main chamber); valving system seeds atoms in Xe or Kr carrier gas before emerging from a pulsed nozzle. **2**) Rotating supersonic source, driven by motor (M) and with pulsed gas inlet valve (PGV), and differentially pumped (DP) feed-in, used for NO$_2$ beam; **3**) solenoid-controlled shutter preceding **4**) skimmer that gives sole entry to detection chamber; **5**) observation zone where parent beams are monitored by a Residual Gas Analyzer (RGA) and laser-induced fluorescence from OH product is recorded by a photomultiplier (PMT).

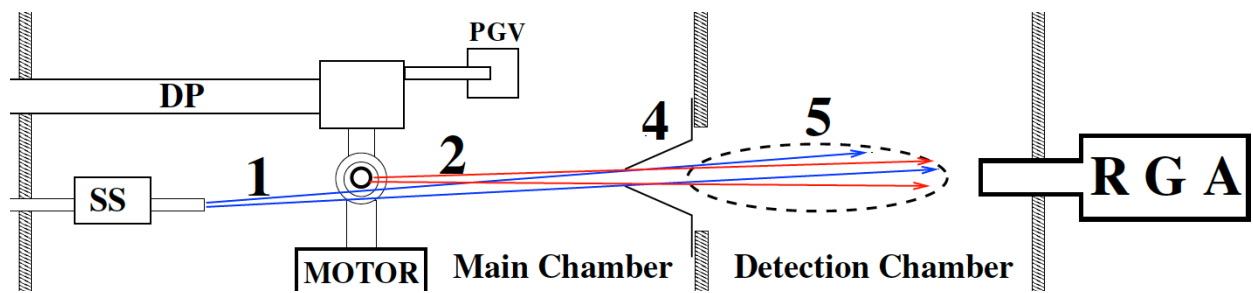

FIG. 2. Schematic (side view, not to scale) showing path of beam **1** (from stationary source, SS) directed at slight angle relative to beam **2** (in the vertical plane) in order to pass below (by ~ 3 mm) the rotor orbit, yet pass through the skimmer **4** (3 mm dia orifice) along with beam **2** so both beams overlap almost completely in the observation zone **5**. The "slight angle" is only about $\arctan(0.3/13) \sim 1.3°$, and the angular beam widths transmitted by the skimmer are about $\theta \sim 1°$. Shutter and PMT are not visible in this projection.

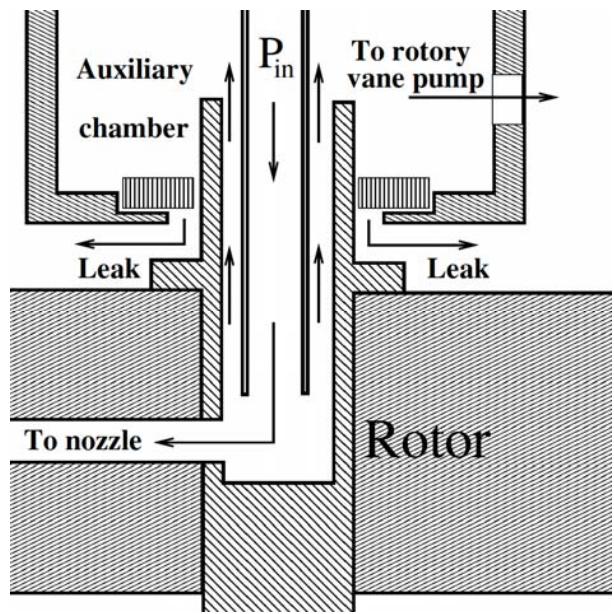

FIG. 3. Schematic (side view) showing how gas is fed from stationary reservoir at pressure $P_{in}$ into spinning rotor (for beam **2** in Figs. 1 and 2). The coupling between the stationary feed tube and the stainless steel inlet to the rotor barrel is housed in an auxiliary chamber that is pumped independently of the main chamber. Cross-section of PEEK washer encircling spigot from rotor is shown as rectangles patterned with vertical lines.

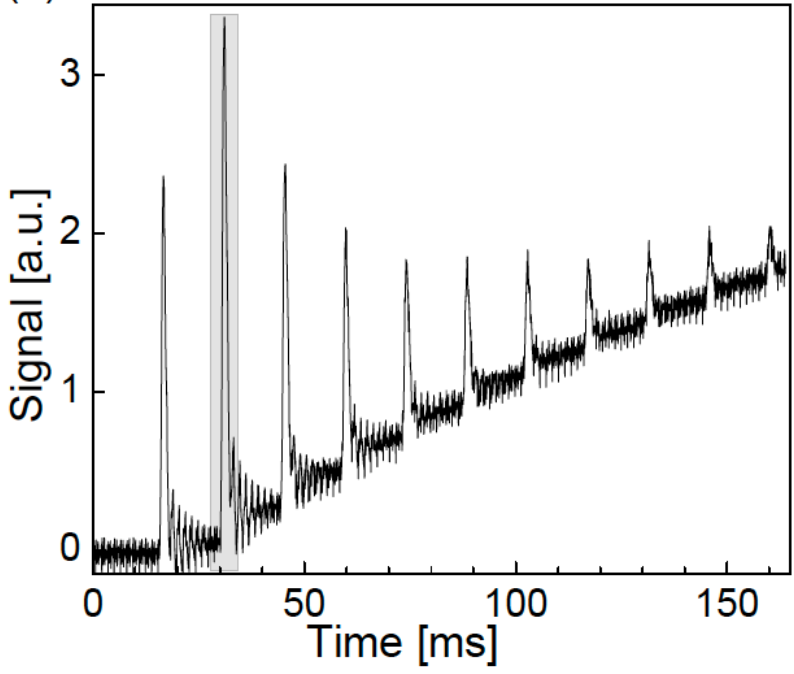

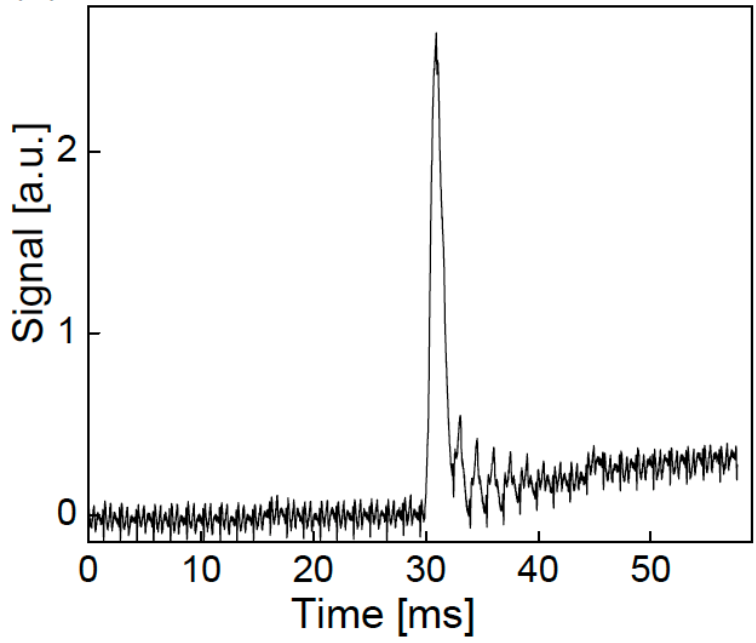

FIG. 4. Raw RGA signals (arbitrary units) vs time (in milliseconds) for pulsed Kr beam, with $P_{in}$ = 300 Torr and rotor spinning at 70 Hz: (a) Sequence of pulses recorded with open shutter; spacing is rotational period of 14 ms, duration of PGV opening was 20 ms. (b) Single pulse [shaded in (a)] was separated from sequence by closing the shutter except during interval from 29 to 33 ms. Growth of background signal in (a) is due to accumulation of gas in detector chamber; closing the shutter suppresses it in (b). The high frequency noise evident in both (a) and (b) is predominantly due to the rectangular pulsed voltage applied to the RGA electron multiplier with frequency 675 Hz; it could be readily filtered out by standard techniques.

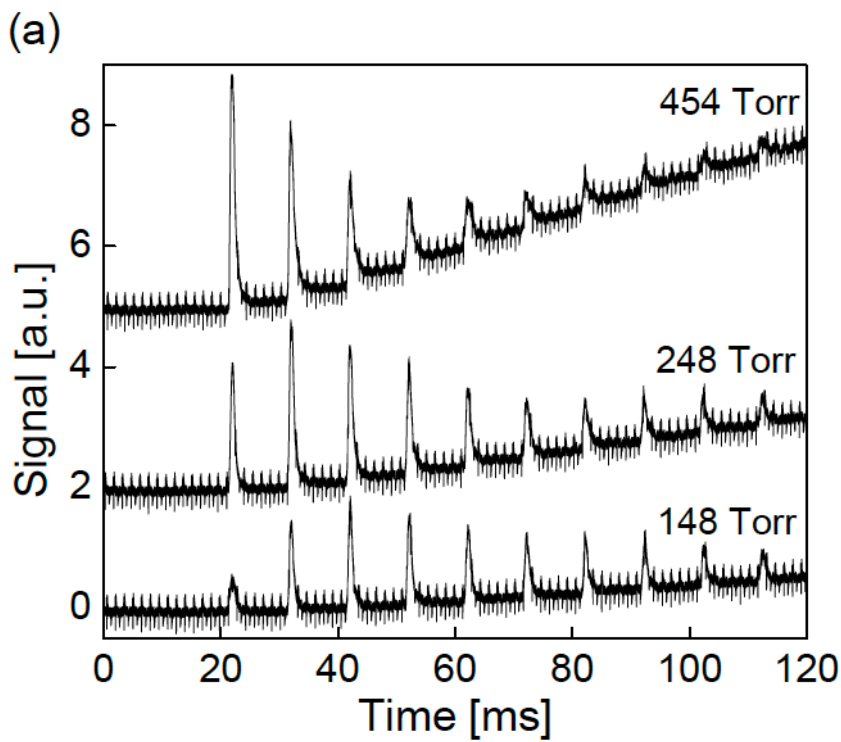

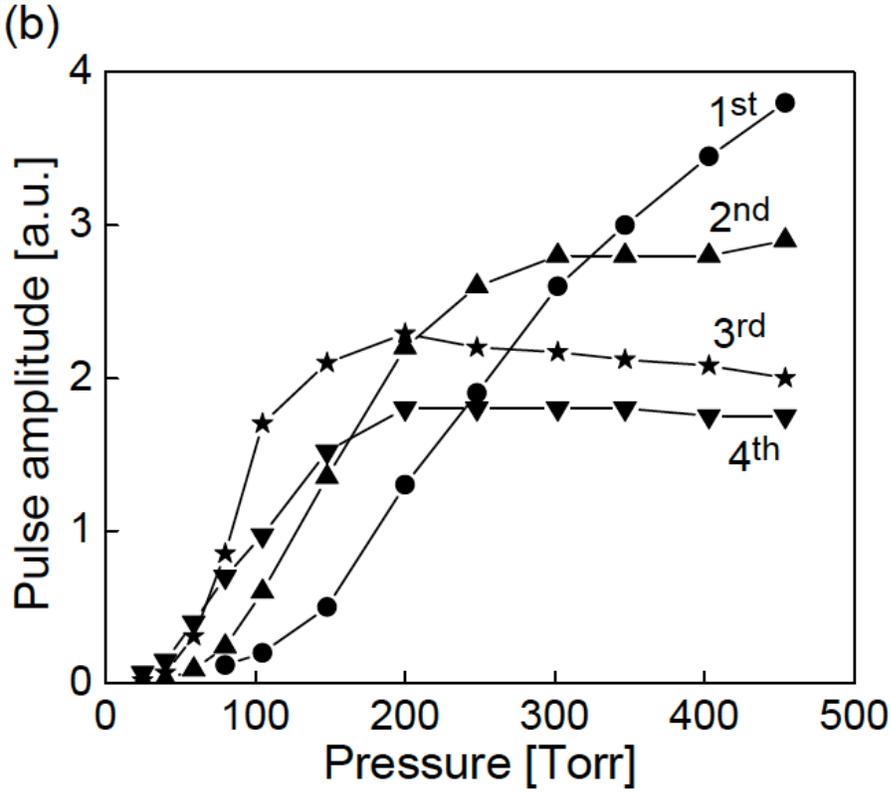

FIG. 5. Comparison of sequences of pulse amplitudes (raw RGA data obtained with open shutter, *cf.* Figure 4) for pulsed Kr beam with rotor spinning at 100 Hz (corresponding to $V_{rot}$ = -94 m/s and $V_{mp}$ ~305 m/s); the open PGV duration was held at 20ms but input pressures ranged from $P_{in}$ = 25 to 454 Torr. (a) Pulses, spaced by $1/\omega$ = 10 ms, for $P_{in}$ = 148, 248, and 454 Torr; for these, the maximum pulse amplitude occurs for the third, second, and first pulse, respectively. (b) Variation of amplitudes of first, second, third, and fourth pulses (at 22, 32, 42, 52 ms, respectively) with input pressure, over range $P_{in}$ = 25 to 454 Torr.

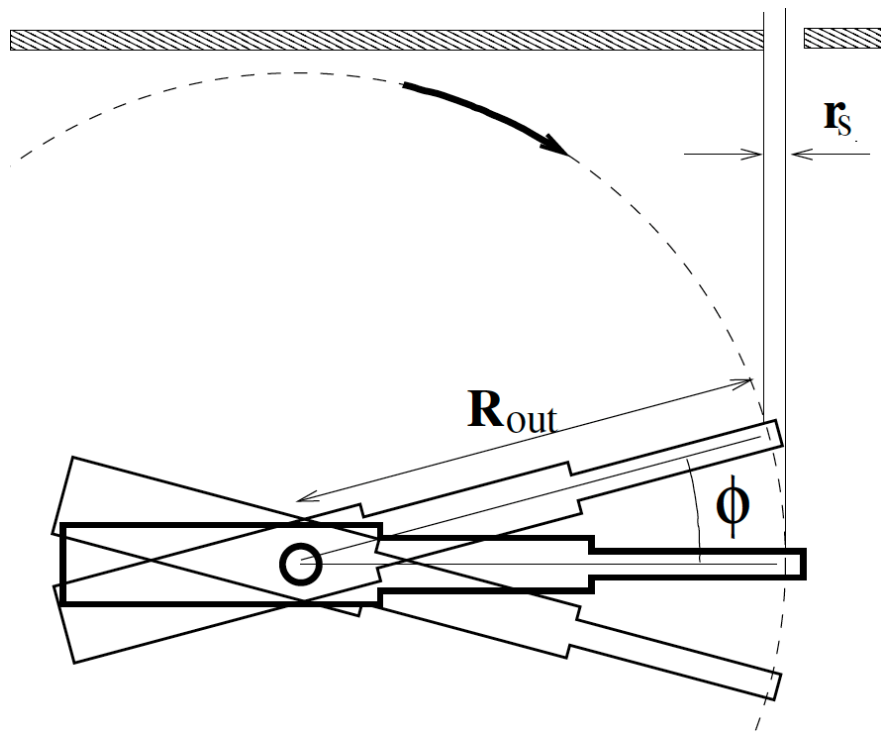

FIG. 6. Spread in "shooting positions" that permit molecules emitted from the rotating source to pass through the skimmer. The maximum angle $\phi$ in the plane of the rotor orbit for which molecules can still enter the skimmer is $\phi_{max} = \cos^{-1}[1 - r_s/R_{out}]$, where $r_s$ is the radius of the skimmer entrance aperture and $R_{out}$ is the length of the rotor barrel between the gas inlet and the exit orifice.

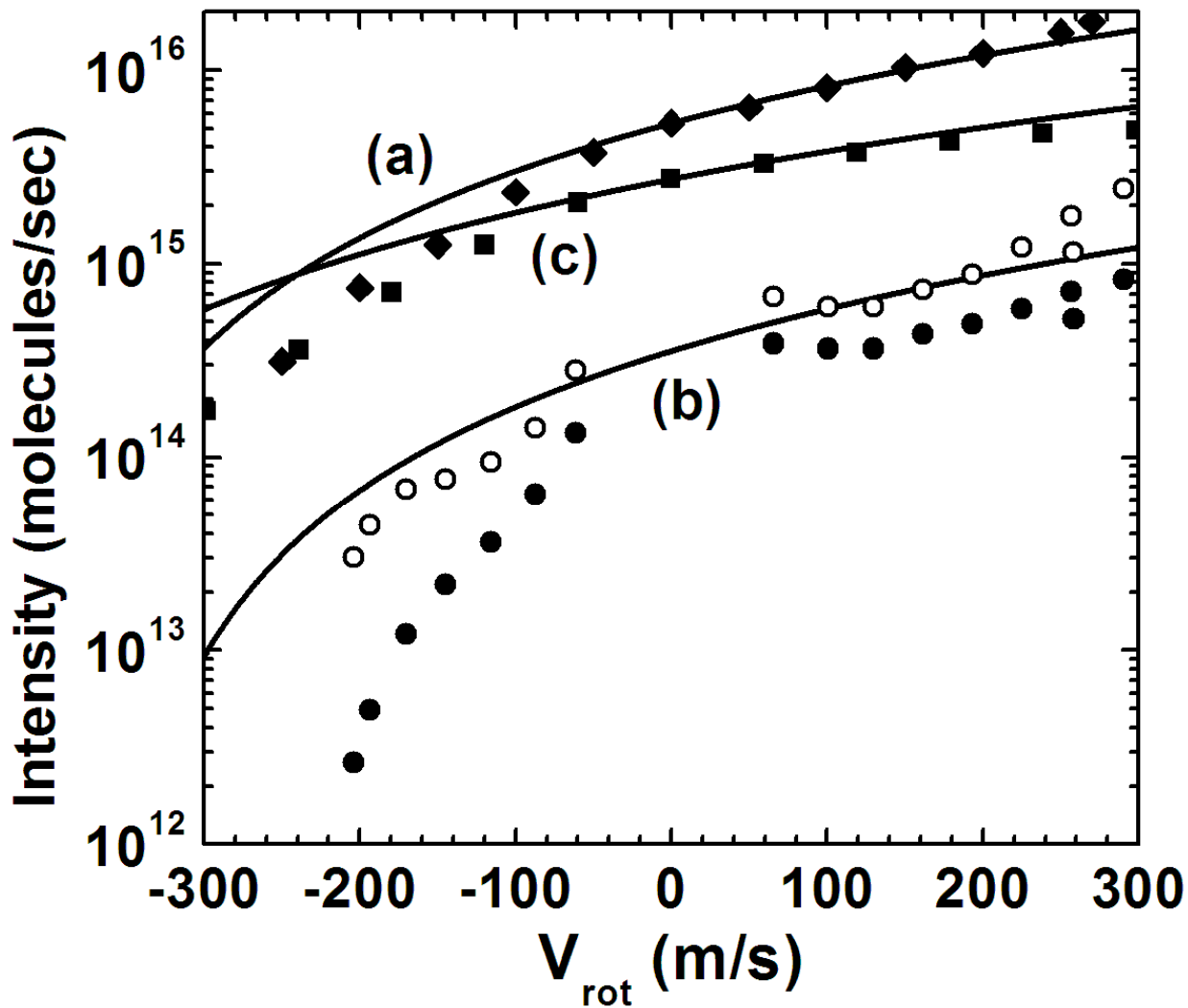

FIG. 7. Variation of intensity with $V_{rot}$ and backing pressure within rotating supersonic nozzle. Curves were calculated from integral of Eq.(2) as described in ref.[35]; the centrifugal contribution of Eq.(1) was not included. Shape of curves depends on the flow velocity u, speed spread $\Delta v/u$ and $V_{rot}/u$; intensity magnitude is proportional to $P_0 d^2$. otherwise involves mostly apparatus factors. Accompanying data points are from: (a) our pulsed source for Kr beam (♦) with $P_{in}$ = 450 Torr; (b) Harvard for Xe beam (•) with $P_{in}$ = 30 Torr; (c) Freiburg for Ar beam (■) with $P_{in}$ = 220 Torr. All three used the same nozzle exit diameter, d = 0.01 cm. For (b) open points (o) have been obtained by correcting experimental data (•) with estimates of attenuation by scattering from background gas. Nominal spread was $\Delta v/u$ = 0.1 and flow velocities used were u = 400, 350, and 550 m/s for (a), (b), (c), respectively.

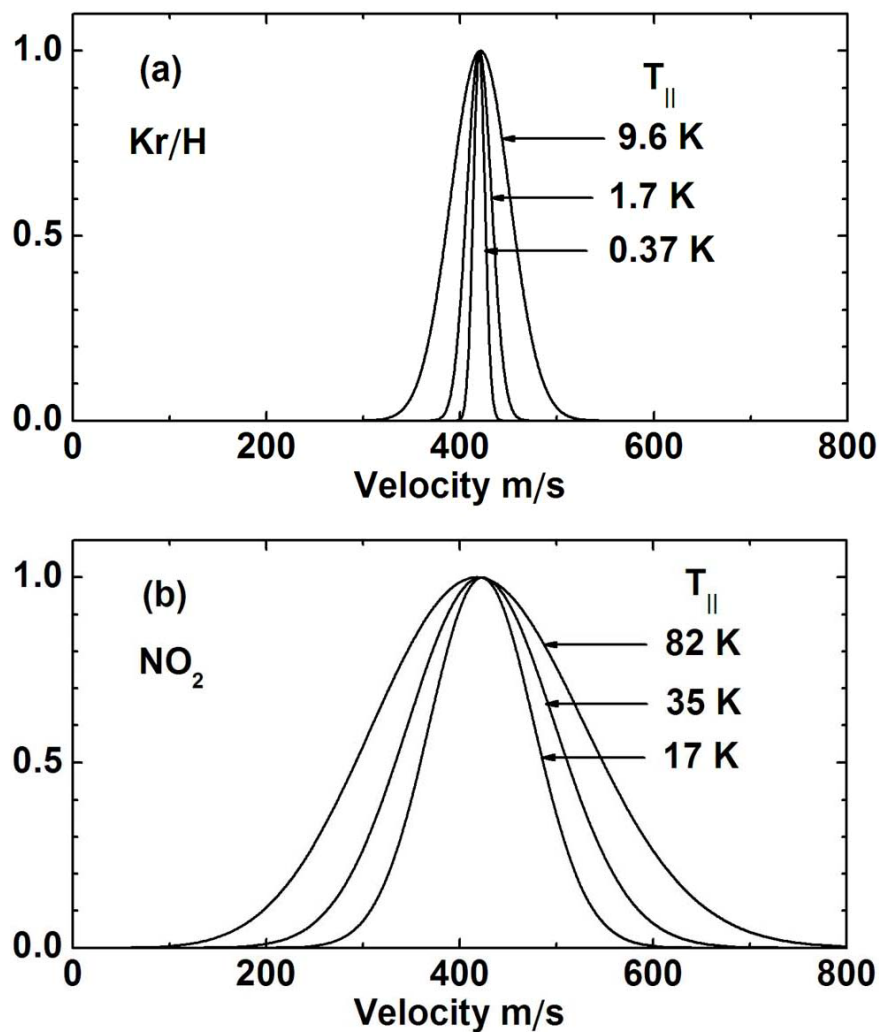

FIG. 8. Velocity distributions for supersonic beams, computed from Eqs.(2-4). Panel (a) pertains to stationary source of beam with a few percent H seeded in Kr; peak position is adjusted upward (from u ~ 385 m/s) to 420 m/s, to account for estimated velocity "slip" of H component. Panel (b) for $NO_2$ beam from counter-rotating source has peak position shifted downwards to 420 m/s (from u ~ 600 – 680 m/s). Longitudinal temperatures $T_{\parallel}$ associate with narrowing of width $\Delta v$ with increase in backing pressure $P_0$ behind nozzle exit orifice. Curves are shown for three values of $P_0 d$ (in Torr-cm); widths obtained from $T_{\parallel}/T_0 = B(P_0 d)^{-\beta}$, with $T_0$ = 300 K; B = 0.0320, $\beta$ = 1.09 for Kr; B = 0.27, $\beta$ = 0.52 for $NO_2$.

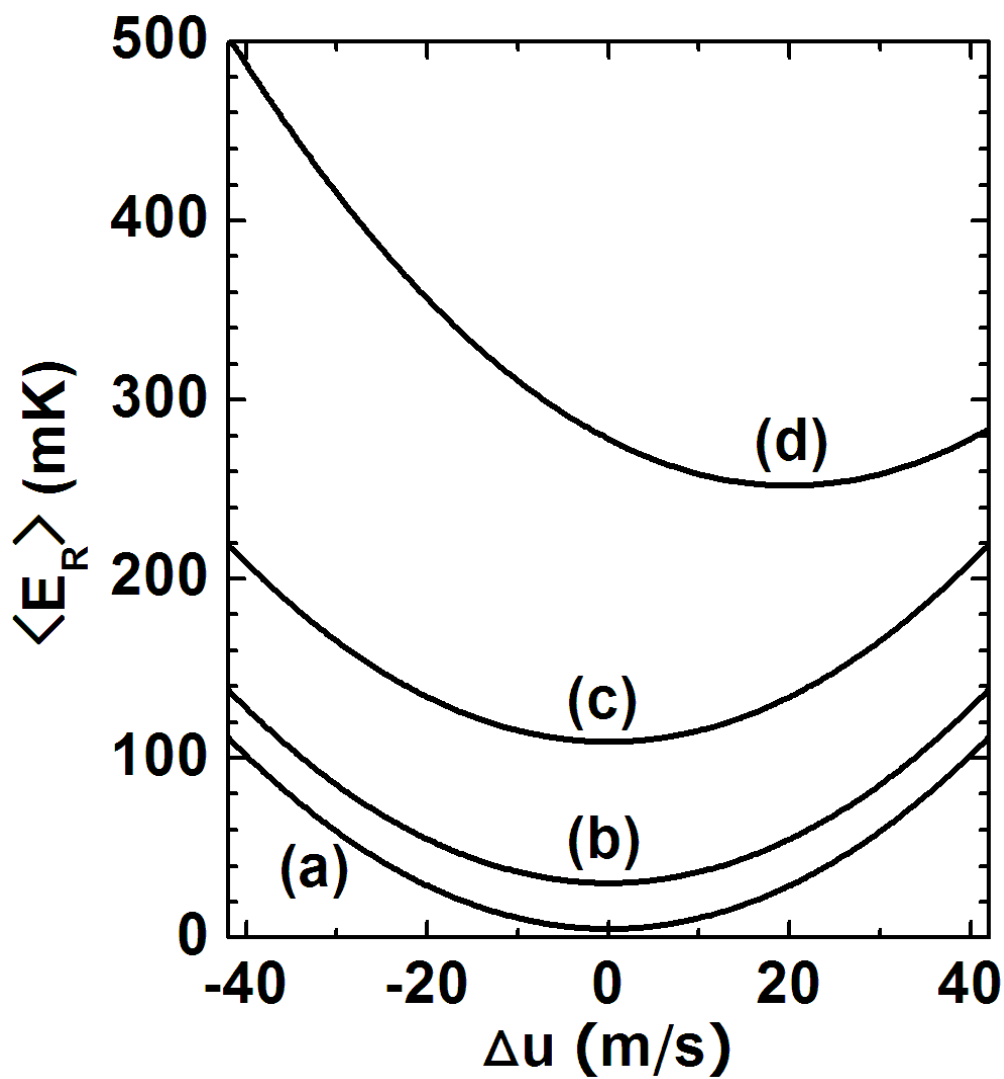

FIG. 9. Relative kinetic energy, $\langle E_R \rangle$, for bimolecular collisions in merged-beams, averaged over velocity distributions; cf. Eq.(6). The plot pertains to supersonic beams with the same most probable flow velocities, $u = u_1 = u_2 = 420$ m/s, reduced mass $\mu = 1$ amu, merging angle spread $\theta = 1°$; curves shown are for various speed spreads, $\Delta v/u$: (a) 1% for both beams; (b) 5% for both beams; (c) 10% for both beams; (d) 10% for one beam, 20% for the other.

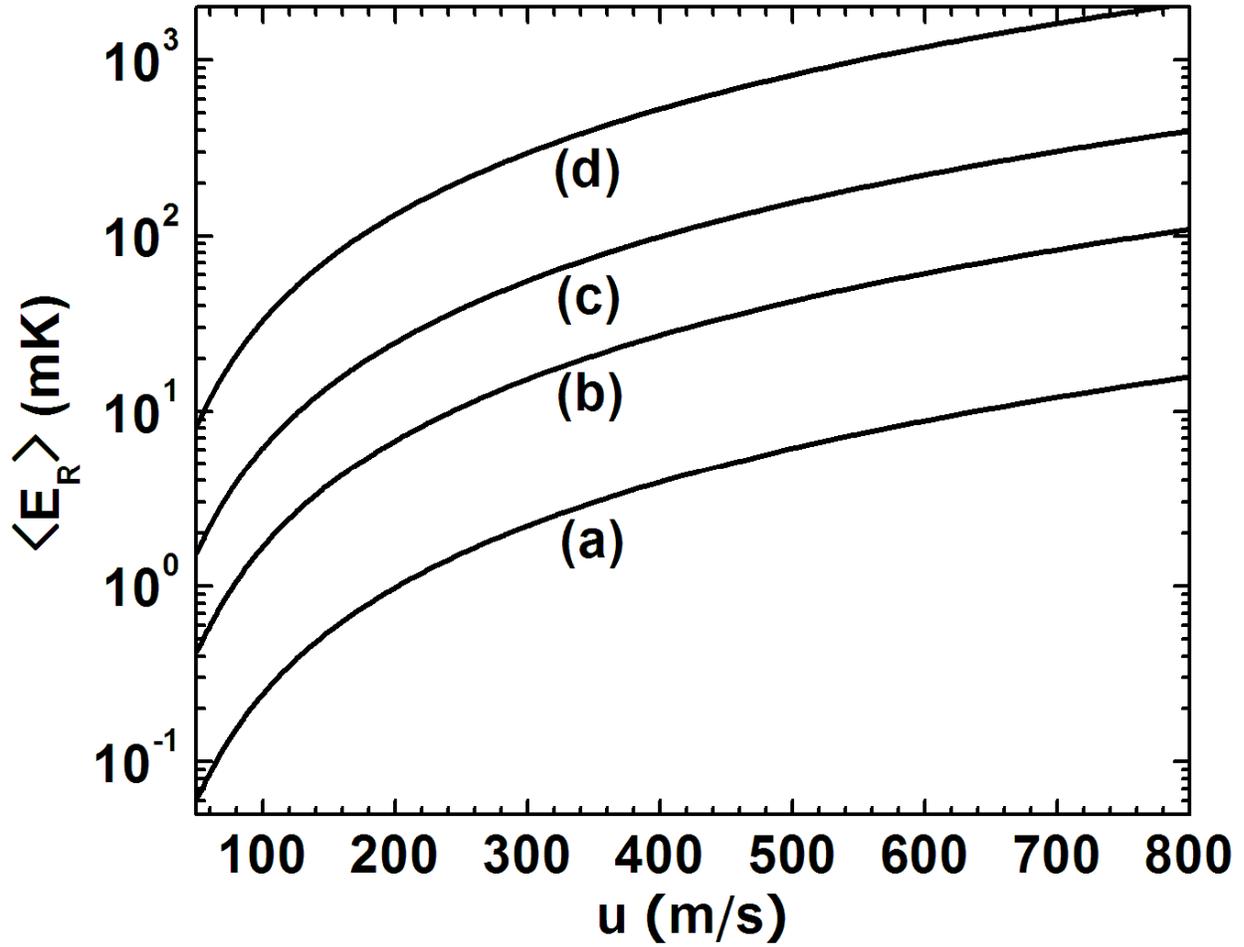

FIG. 10. Relative kinetic energy, $\langle E_R \rangle$, as specified in Figure 9 but for beams with most probable flow velocities that may differ, $\Delta u = u_1 - u_2$, within the range $420 \pm 40$ m/s. Again, curves shown are for various speed spreads, $\Delta v/u$; in (d) beam **1** has 10% spread, beam **2** has 20%.